\documentclass[aps,showpacs,preprintnumbers,amsmath,amssymb,showpacs,showkeys]{revtex4}
\usepackage{amssymb}
\usepackage{amsmath}
\usepackage{amsfonts}
\usepackage{epsfig}
\usepackage{graphicx}
\usepackage{tabularx}
\usepackage{latexsym,epsfig,amsmath,amssymb}
\usepackage{graphics}
\usepackage{subfigure}
\usepackage{verbatim}
\newcommand{\seq}{\begin{subequations}}
\newcommand{\sen}{\end{subequations}}
\newcommand{\eq}{\begin{eqnarray}}
\newcommand{\en}{\end{eqnarray}}
\newcommand{\ra}{\rangle}

\def\L2{\Lambda^2}

\def\eq{\begin{eqnarray}}
\def\en{\end{eqnarray}}
\def\zb{Z_b^+}
\def\zbp{Z_b^{'+}}

\def\L2{\Lambda^2}

\def\eq{\begin{eqnarray}}
\def\en{\end{eqnarray}}
\def\zc{Z_c^+}
\def\zcp{Z_c^{'+}}
\def\zb{Z_b^+}
\def\zbp{Z_b^{'+}}

\begin{document}

\title{Strong decays of molecular states $\zc$ and $\zcp$}

\author{Yubing Dong$^{1,2}$,
        Amand  Faessler$^3$,
        Thomas Gutsche$^3$,
        Valery E. Lyubovitskij$^3$\footnote{On leave of absence
        from Department of Physics, Tomsk State University,
        634050 Tomsk, Russia}
\vspace*{1.2\baselineskip}}
\affiliation{
$^1$ Institute of High Energy Physics, Beijing 100049, P. R. China\\
\vspace*{.3\baselineskip}\\
$^2$ Theoretical Physics Center for Science Facilities (TPCSF), CAS,
Beijing 100049, P. R. China\\
\vspace*{.3\baselineskip}\\
$^3$ Institut f\"ur~Theoretische Physik, Universit\"at T\"ubingen,\\
Kepler Center for Astro and Particle Physics, \\
Auf der Morgenstelle 14, D--72076 T\"ubingen, Germany\\}

\date{\today}

\begin{abstract}

The newly observed hidden-charm meson $\zc(3900)$ and a possible partner 
state $\zcp$ with quantum numbers $J^P = 1^+$ are considered as hadronic 
molecules composed of $\bar{D}D^*$ and $\bar{D}^*D^*$, respectively. 
We give predictions for the decay widths of the strong two-body transitions
$\zc\to H+\pi^+$ and $\zcp\to H+\pi^+$ with $H = \Psi(nS), h_c(mP)$ 
in a phenomenological Lagrangian approach.

\end{abstract}



\maketitle

\section{Introduction}

Recently the three collaborations
BESIII~\cite{Ablikim:2013mio}, Belle~\cite{Liu:2013dau} and
CLEO-c~\cite{Xiao:2013iha} reported about the observation of a new resonance
$Z_c(3900)$ with a mass $(3899\pm 3.6\pm 4)$~MeV and a width of
$(46\pm 10\pm 20)$~MeV~\cite{Ablikim:2013mio}.
The observation of this state already motivated a series of theoretical studies
based on different
assumptions (mainly hadronic molecular and tetraquark interpretations were
discussed).
Here we analyze the strong two-body decays of
$\zc$ and its possible partner state $\zcp$
using a phenomenological Lagrangian
approach~\cite{Faessler:2007gv}-\cite{Dong:2012hc} based on 
the compositeness condition~\cite{Weinberg:1962hj}-\cite{Branz:2009cd}, 
which was successfully applied for the study of hadrons and exotic 
states as bound states of their constituents using methods of 
quantum field theory. 

The main idea of 
the compositeness condition~\cite{Weinberg:1962hj}-\cite{Branz:2009cd} is 
to define the coupling strength of the field representing 
the bound state and their constituents from the equation 
$Z=0$~\cite{Weinberg:1962hj,Efimov:1993ei}. Here 
$Z$ is the wave function renormalization constant of the field describing 
the bound state. The quantity $Z^{1/2}$ is the matrix element between 
a physical particle state and the corresponding bare state. 
The compositeness condition $Z=0$ enables 
one to represent a bound state by introducing a hadronic field interacting 
with its constituents so that the renormalization factor is equal to zero. 
This does not mean that we can solve the QCD bound state equations but we are 
able to show that the condition $Z=0$ provides an effective and 
self--consistent way to describe the coupling of a hadron to its 
constituents. One  starts with an phenomenological interaction Lagrangian 
written down in terms of the field describing bound states and their 
constituents. Then, by using Feynman rules, 
the $S$--matrix elements describing hadron-hadron interactions are given in 
terms of Feynman loop diagrams with constituents running in the loops. 
The compositeness condition enables one to avoid the problem 
of double counting. The approach is self--consistent and all calculations 
of physical observables are straightforward. There is  a small set of 
model parameters: the values of the constituent masses and the scale 
parameters that define the size of the distribution of the constituents 
inside a given bound state. 

We consider the $\zc$ state as 
a hadronic molecule as also discussed 
previously and extensively among the theoretical interpretations
collected in~Refs.~\cite{Guo:2013sya}. In addition we extend the considerations
to a possible partner state $\zcp$. In particular, we treat
the charged hidden-charm meson resonances $\zc$ and $\zcp$ as a
superposition of the molecular configurations $\bar{D}D^{*}$ and
$\bar{D}^*D^*$ as
\eq
|Z_c^+\ra &=&\frac{1}{\sqrt{2}} \Big| D^{*+}\bar{D}^0+\bar{D}^{*0}D^
+ \Big\ra,
\nonumber \\
|Z_c^{+'}\ra &=& |D^{*+}\bar{D}^{*0}\ra.
\en
We adopt the spin and parity quantum numbers
$J^P = 1^{+}$ for the two resonances $\zc$ and $\zcp$. 
Note the bottomia states ($\zb$ and $\zbp$) have been considered in our 
approach in Ref.~\cite{Dong:2012hc}. 

In the present paper we proceed as follows. In Sec.~II we briefly review the
basic ideas of our approach where we set up the two new resonances
$\zc$ and $\zcp$ as $\bar{D}D^*$ and
$\bar{D}^*D^*$ molecular states. 
Then we proceed to consider the strong two-body
decays $\zc(\zcp) \to \Psi(nS)+\pi^+$ and $\zc(\zcp) \to h_c(mP)+\pi^+$
based on a phenomenological interaction Lagrangian.
In Sec.~III we present the numerical results and discussion.

\section{Framework}

Our approach to the  $\zc$ and $\zcp$ states is based on interaction
Lagrangians describing the coupling of the $\zc$ and $\zcp$
states to its constituents as
\eq\label{Lagr}
{\cal L}_{Z_c}(x)&=&\frac{g_{_{Z_c}}}{\sqrt{2}}\, M_{Z_c} \,
Z_c^{\mu}(x)\int d^4y \, \Phi_{Z_c}(y^2)\Big (D(x+y/2)
\bar{D}^*_{\mu}(x-y/2)+
D^*_{\mu}(x+y/2)\bar{D}(x-y/2)\Big ), \nonumber\\
{\cal L}_{Z_c'}(x)&=&\frac{g_{_{Z_c'}}}{\sqrt{2}} \,
i \epsilon_{\mu\nu\alpha\beta}
\partial^{\mu}Z_c^{'\nu}(x) \int d^4y \,
\Phi_{Z_c'}(y^2)D^{*\alpha}(x+y/2)\bar{D}^{*\beta}(x-y/2),
\en
where $y$ is the relative Jacobi coordinate (difference of coordinates of the 
constituents), $g_{_{Z_c}}$ and
$g_{_{Z_c^\prime}}$ are the dimensionless coupling constants of $\zc$ and
$\zcp$ to the molecular $\bar{D}D^{*}$ and $\bar{D}^*D^{*}$
components, respectively. Here $\Phi_{Z_c}(y^2)$ and $\Phi_{Z_c^\prime}(y^2)$
are correlation functions, which describe the distributions of
the constituent mesons in the bound states. A basic requirement for the
choice of an explicit form of the correlation function $\Phi_H(y^2)$
($H = Z_c, Z_c^\prime$) is that its Fourier transform vanishes sufficiently
fast in the ultraviolet region of Euclidean space to render the Feynman
diagrams ultraviolet finite. We adopt a Gaussian form for the correlation
function. The Fourier transform of this vertex function is given by
\eq
\label{corr_fun}
\tilde\Phi_H(p_E^2/\Lambda^2) \doteq \exp( - p_E^2/\Lambda^2)\,,
\en
where $p_{E}$ is the Euclidean Jacobi momentum.
$\Lambda$ is a size parameter characterizing the distribution of the two
constituent mesons in the $\zc$ and $\zcp$ systems, which also leads to a
regularization of the ultraviolet divergences in the Feynman diagrams.
For a molecular system where the binding energy is negligible in
comparison with the masses of the constituents this size parameter is
expected to be smaller than 1 GeV. 
From our previous analyses of the strong two-body decays of the
$X, Y, Z$ meson resonances and of the $\Lambda_c(2940)$ and $\Sigma_c(2880)$
baryon states we deduced a value of maximally
$\Lambda \sim 1$~GeV~\cite{Dong:2009tg}.
For a very loosely bound system like the $X(3872)$ a size parameter of
$\Lambda \sim 0.5$~GeV~\cite {Dong:2009uf} is more suitable.
For heavy compact states such as tetraquark states, charmonia 
or a possible charmonium component in
the X(3872) the size parameter $\Lambda $ is typically much larger
(for example in a range from 2.5 to 3.5 GeV as discussed in 
Ref.~\cite {Dong:2009uf}).
Here we choose values for $\Lambda $ in the range $0.5-0.75$ GeV which
reflect a weakly bound heavy meson system.
Once $\Lambda $ is fixed the coupling constants $g_{_{Z_c}}$ 
and $g_{_{Z_c^\prime}}$ are then determined by the  compositeness 
condition~\cite{Faessler:2007gv}-\cite{Branz:2009cd}. It implies that the 
renormalization constant of the hadron wave function is set equal to zero with:
\eq\label{ZLc}
Z_H  = 1 - \Sigma_H^\prime(M_H^2) = 0 \,.
\en
Here, $\Sigma_H^\prime$ is the derivative of the transverse part of the
mass operator $\Sigma_H^{\mu\nu}$ of the molecular states (see Fig.1),
which is defined as
\eq
\Sigma_H^{\mu\nu}(p) = g^{\mu\nu}_\perp \, \Sigma_H(p)
+ \frac{p^\mu p^\nu}{p^2}
\Sigma_H^L(p)\,, \quad
g^{\mu\nu}_\perp = g^{\mu\nu} - \frac{p^\mu p^\nu}{p^2} \,.
\en
Analytical expressions for the couplings
$g_{_{Z_c}}$ and $g_{_{Z_c^\prime}}$ are given in Appendix A.
In the calculation the masses of $Z_c$ and $Z_c^\prime$
are expressed in terms of the constituent masses and the binding
energy $\epsilon$ (a variable quantity in our calculations):
\eq
M_{Z_c}=M_{D}+M_{D^*}-\epsilon\,, \quad
M_{Z_c^\prime}=2M_{D^*}-\epsilon\,,
\en
where $\epsilon$ is the binding energy.

In the calculation of the two-body decays of
$\zc(\zcp)\to H + \pi^+$ where $H = \Psi(nS), h_c(mP)$
we generate the four-particle $DD^\ast H \pi^+$ and
$D^*D^\ast H \pi^+$ vertices by a phenomenological Lagrangian
\eq\label{Lagr_eff}
{\cal L}_{\cal D \bar D H P}(x) =
i g_F \,{\rm tr} \Big( \bar{\cal D}(x) \,
[ {\cal H}(x), {\cal P}(x)] \, {\cal D}(x) \Big)
\, + \,
g_D \,{\rm tr} \Big( \bar{\cal D}(x) \,
\{ {\cal H}(x), {\cal P}(x)\} \, {\cal D}(x) \Big)
\,,
\en
where $g_F$ and $g_D$ are effective coupling constants,
$[\ldots]$ and $\{\ldots\}$ denote the commutator and
anticommutator, respectively.

The $H$ is the heavy charmonia field; $D$ is the superposition of
isodoublets of open-charm mesons with $J^P = 0^-, 1^-$ and $1^+$;
${\cal P}$ is the chiral field:
\eq
{\cal H} &=& J^\mu \gamma_\mu + h^\mu \gamma_\mu \gamma_5
\, + \, \frac{g_{H}}{M_H} \,
\Big( J^{\mu\nu} \, \sigma_{\mu\nu} \, +  \,
h^{\mu\nu} \sigma_{\mu\nu} \gamma_5 \Big) \,, \\
{\cal D} &=& D i\gamma_5 + D^{\ast\,\mu} \gamma_\mu
          +  D_1^{\mu} \gamma_\mu \gamma_5  \,,\\
{\cal P} &=& \frac{1}{2} \not\! u \, \gamma^5
          + \frac{1}{2} [u^\dagger,\partial_\mu u] \gamma^\mu \,,
\en
where $J$ and $h$ denote the $\Psi$ and $h_c$ states; 
$V^{\mu\nu} = \partial^\mu V^\nu - \partial^\nu V^\mu$ is 
the stress tensor of the $\Psi$ and $h_c$ states;
$g_{\cal H}$ is a  phenomenological
coupling  defining the mixing of derivative and nonderivative terms
in ${\cal H}$; $M_H \simeq M_J$ is associated with the $J/\psi$ mass;
$D = (D^+, D^0)$, $D^\ast = (D^{\ast\, +}, D^{\ast\, 0})$
are the doublets of pseudoscalar and vector
charmed $D$ mesons; $u_\mu$ is the chiral vielbein:
\eq
u_\mu = i \{ u^\dagger, \partial_\mu u \}\,, \quad u^2 = U
= \exp\Big[i\frac{\hat{\pi}}{F_\pi}\Big]\,, \quad
\hat{\pi} = \vec{\pi} \vec{\tau}
\en
where $F_\pi = 92.4$ MeV is the pion decay constant,
$\vec{\pi}=(\pi_1,\pi_2,\pi_3)$ is the triplet of pions.
Note the couplings $g_D$, $g_F$ and $g_H$ are phenomenological
parameters. Below we show that we have two constraints on these couplings.

From Eq.~(\ref{Lagr_eff}) we deduce specific Lagrangians describing the
couplings between heavy charmonia, charmed mesons and the pion which
are relevant for the decays of the $Z_c$ and $Z_c^\prime$ states:
\eq\label{Lagr_phen}
{\cal L}_{DD^\ast J\pi}(x) &=&-\frac{8g_Fg_H}{F_\pi M_J} \,
J^{\mu\nu}(x) \, \bar D^\ast_\nu(x) \,
\partial_\mu \hat{\pi}(x) \,
D(x) \, + \, {\rm H.c.} \,, \\
{\cal L}_{D^\ast D^\ast J\pi}(x) &=&\frac{4g_D}{F_\pi} \,
\varepsilon^{\mu\nu\alpha\beta} \, J_{\mu} \,
\bar D^{\ast}_\beta(x) \,
i \partial_\nu \hat{\pi}(x) \,
D^{\ast}_\alpha(x) \,, \\
{\cal L}_{DD^\ast h_c\pi}(x) &=&-\frac{4g_Fg_H}{F_\pi M_J} \,
\varepsilon^{\mu\nu\alpha\beta} \
h_{\mu\nu}(x) \, \bar D^{\ast}_\alpha(x) \,
 \partial_\beta \hat{\pi}(x) \, D(x)
\, + \, {\rm H.c.} \,, \\
{\cal L}_{D^\ast D^\ast h_c \pi}(x) &=& \frac{4g_F}{F_\pi} \,
\bar D^{\ast\,\nu}(x) \,
  ( h_\mu(x) \, i\partial_\nu \hat{\pi}(x)
-   h_\nu(x) \, i\partial_\mu \hat{\pi}(x) )
\,  D^{\ast\,\mu}(x) \,.
\en
The three-particle coupling $g_{D^\ast D\pi}$ of the pion to charmed 
$D$ mesons is defined by the phenomenological
Lagrangian:
\eq
{\cal L}_{D^\ast D \pi}(x)&=&\frac{g_{D^\ast D \pi}}{\sqrt{2}} \,
\bar D^{\ast\,\mu}(x) \partial_\mu \hat{\pi}(x) D(x) + \, {\rm H.c.} \,,
\en
where the value $g_{D^\ast D \pi} = 17.9$ has been determined from data
on $D^\ast \to D \pi$ decay~\cite{Anastassov:2001cw}. 
The coupling $D^\ast D \pi$  has been calculated e.g. using QCD sum rules,  
first in Ref.~\cite{Belyaev:1994zk} and it was updated in several papers. 
One of the latest estimates is given in Ref.\cite{Navarra:2001ju}.  
The first estimate of this coupling from lattice QCD was done in 
Ref.~\cite{Abada:2002vj} and updated 
in Ref.~\cite{Becirevic:2009xp}. 

Next we discuss how we fix the couplings $g_D$, $g_F$ and $g_H$.
As mentioned before these parameters can be further constrained.
In particular, we can relate the coupling $g_D$ and the product
$g_F g_H$ to the four-particle couplings $g_{D^\ast D J\pi}$
and $g_{D^\ast D J\pi}$ which appear in the phenomenological Lagrangian
proposed in Ref.~\cite{Lin:1999ad} for the analysis of $J/\psi$ absorption
in hadronic matter (see details in Appendix B). 
Note, that coupling of $D$ mesons with pion and $J/\Psi$ were calculated  
also in Ref.~\cite{Deandrea:2003pv}.  

Matching of the coupling constants leads to
\eq
g_{D^\ast D J \pi} &=&
\frac{g_{JDD} \ g_{D^\ast D\pi}}{2 \, \sqrt{2}}  \simeq
\frac{8 g_F g_H }{F_\pi M_J \sqrt{6}} \, (M_{Z_c}^2-M_J^2) \,
\biggl( 1 + \frac{M_J^2}{2M_{Z_c}^2} \biggr) \,, \nonumber\\
g_{D^\ast D^\ast J \pi} &=& \frac{2 g_D}{F_{\pi}} \,
\frac{M_{Z_c'}^2-M_J^2}{M_{Z_c'}^2}\, .
\en
The four-particle couplings $g_{D^\ast D^\ast J \pi}$ and
$g_{D^\ast D J \pi}$ were expressed in Ref.~\cite{Lin:1999ad}
in a factorization of two three-particle couplings
\eq
g_{D^\ast D J \pi} \, = \, 2 \sqrt{M_D M_{D^\ast}} \, g_{D^\ast D^\ast J \pi}
\, = \, \frac{g_{JDD} g_{D^\ast D\pi}}{2 \sqrt{2}} \,.
\en
Therefore, in the numerical evaluation we use an approximate 
condition $g_D \simeq g_F$ 
and $g_H$ is fixed by the condition (following the discussion above)
\eq
g_H \simeq  g_{JDD} \ g_{D^\ast D\pi} \,
\frac{F_\pi M_J \sqrt{3}}{16 \, g_F \, (M_{Z_c}^2-M_J^2)} \,
\biggl( 1 + \frac{M_J^2}{2M_{Z_c}^2} \biggr)^{-1} \,.
\en

\section{Numerical results}

With the phenomenological Lagrangians introduced and discussed
we can proceed to determine the widths of the two-body decays
$\zc(\zcp) \to \Psi(nS)+\pi^+$ and
$\zc(\zcp) \to h_c(mP)+\pi^+$. The relevant diagrams are
indicated in Fig.2. The standard evaluation leads to the
corresponding decay widths:
\eq
\Gamma_{\zc\to \Psi(nS) \pi^+}&\simeq&
 \frac{g_{Z_c\Psi(nS)\pi}^2}{96\pi M_{Z_c}^3}
\lambda^{3/2}(M_{Z_c}^2, M_{\Psi(nS)}^2, M_{\pi}^2) \,
\biggl(1 + \frac{M_{\Psi(nS)}^2}{2M_{Z_c}^2}\biggr)\,, \nonumber\\
\Gamma_{\zcp\to\Psi(nS)\pi^+}&\simeq&
\frac{g_{Z_c^\prime\Psi(nS)\pi}^2}{96 \pi M_{Z_c^\prime}^3}
\lambda^{3/2}(M_{Z_c^\prime}^2, M_{\Psi(nS)}^2, M_{\pi}^2) \,
\biggl(1 + \frac{M_{Z_c'}^2-M_{\Psi(nS)}^2}{3M_{\Psi(nS)}^2}\biggr)\,,
\nonumber\\
&&\\
\Gamma_{\zc\to h_c(mP) \pi^+}&\simeq&
 \frac{g_{Z_ch_c(mP)\pi}^2}{96\pi M_{Z_c}^3}
\lambda^{3/2}(M_{Z_c}^2, M_{h_c}^2, M_{\pi}^2) \,
\biggl(1 + \frac{M_{h_c}^2}{2M_{Z_c}^2}\biggr)\,, \nonumber\\
\Gamma_{\zcp\to h_c(mP) \pi^+}&\simeq&
 \frac{g_{Z_c'h_c\pi}^2}{96\pi M_{Z_c'}^3}
\lambda^{3/2}(M_{Z_c'}^2, M_{h_c}^2, M_{\pi}^2) \,, \nonumber
\en
where $\lambda(x,y,z)=x^2+y^2+z^2-2xy-2xz-2yz$
is the K\"allen function.
The decay coupling constants
$g_{Z_c\Psi(nS)\pi}$,
$g_{Z_c'\Psi(nS)\pi}$,
$g_{Z_ch_c(mP)\pi}$ and
$g_{Z_c'h_c(mP)\pi}$ are expressed by
\eq\label{couplings}
g_{Z_c\Psi(nS)\pi} &=& g_{Z_ch_c(mP)\pi} \, = \,
8 \, g_{Z_c} \frac{g_F g_H}{F_\pi M_J} \, J_1 \, M_{Z_c}
\,, \nonumber\\
g_{Z_c'\Psi(nS)\pi} &=& g_{Z_c'h_c(mP)\pi} \frac{g_D}{g_F} \,
\sqrt{\frac{3}{2}} \, = \, 4 \sqrt{\frac{3}{2}} \,
g_{Z_c'} \frac{g_D}{F_\pi} \, J_2 \,,
\en
where $g_{Z_c}$, $g_{Z_c'}$ and the loop integrals $J_1$ and $J_2$ are given in
Appendix A. We present our results in Tables I-III.
In Table I we display the predictions for the phenomenological
couplings $g_{_{Z_cH\pi}}$ and
$g_{_{Z_c'H\pi}}$ as defined in Eq.~(\ref{couplings})
for different values of the binding energy 
$\epsilon$ and $\Lambda$ varied from 0.5 to 0.75 GeV. 
For convenience, in Table II we present the values for the K\"allen
functions in order to explain the results for the widths given
in Table III. The decay rates for the $\zcp$ states are larger than the
corresponding ones of the $\zc$ resonances, the decay hierarchies
with $\Gamma (1S)>\Gamma (2S)>\Gamma (1P)$ are identical for both states.
An increase of the size parameter $\Lambda$, more suitable for a compact
bound state, would also lead to a sizable increase in the decay rates.
Therefore, if experiment will deliver  larger values
for the rates than predicted in our approach, it would signal that
$\zc$ and $\zcp$ are probably not molecular states.

In summary, using a phenomenological Lagrangian approach we give predictions
for the two-body decay rates of the $\zc$ and $\zcp$ states interpreted
as hadronic molecules. The results could be useful for forthcoming
measurements of these decay modes. 
In future we plan to consider radiative and other strong decays
of the $\zc$ and $\zcp$ mesons.

\begin{acknowledgments}

This work is supported by the DFG under Contract No. LY 114/2-1,
National Sciences Foundations of China No.10975146 and 11035006,
and by the DFG and the NSFC through funds provided to the sino-German CRC 110
``Symmetries and the Emergence of Structure in QCD''.
The work is done partially under the project 2.3684.2011 of Tomsk State
University. One of us (YBD) thanks the Institute of Theoretical Physics,
University of T\"ubingen for the warm hospitality and thanks the support
from the Alexander von Humboldt Foundation. Discussions with Dr. D. Y. Chen
are appreciated.

\end{acknowledgments}

\appendix

\appendix

\section{Matching of the coupling constants $g_D$, $g_F$ and $g_H$}

The idea for matching the coupling constants $g_D$, $g_F$ and $g_H$ is
based on the equivalence of matrix elements squared (or decay
rates) calculated in different approaches --- 
from one side using phenomenological Lagrangians~(\ref{Lagr_phen})
and from other side the Lagrangians
proposed in Ref.~\cite{Lin:1999ad} 
\eq
{\cal L}_{DD^\ast J\pi}(x)&=&g_{_{D^\ast D J\pi}}J_{\mu}(x)
\bar{D}^{\ast\mu}(x)\vec{\pi}(x)\cdot\vec{\tau}D(x) + \mathrm{H.c.}\,,
\label{DDSJPpi}\\
{\cal L}_{D^\ast D^\ast J\pi}(x)&=&i\epsilon_{\mu\nu\alpha\beta}
g_{_{D^\ast D^\ast J\pi}} \, \Big (J^{\mu}(x)
\bar{D}^{\ast\beta}(x)\partial^{\nu}\vec{\pi}(x)
\cdot\vec{\tau}D^{\ast\alpha}(x)\nonumber\\
&+&\partial^{\nu}J^{\mu}(x)
\bar{D}^{\ast\beta}(x)\vec{\pi}(x)
\cdot\vec{\tau}B^{\ast\alpha}(x)\Big ) \,.
\label{DSDSJpi}
\en
Evaluating the matrix elements squared and averaging over
the polarizations of the particle spins we get
in case of the $Z_c \to J/\psi + \pi$ transition using the Lagrangian
of Ref.~\cite{Lin:1999ad}
\eq\label{Matr1_Ko}
\sum\limits_{\rm pol} |M_{\rm inv}|^2 =
g_{D^\ast D J \pi}^2 \,  \biggl(3 - \frac{M_\pi^2}{M_Z^2}
+ \frac{(p_1p_2)^2}{M_Z^2 M_J^2} \biggr) = g_{D^\ast D J \pi}^2 \,
\biggl( 3 + \frac{\lambda(M_Z^2,M_J^2,M_\pi^2)^2}{4 M_Z^2 M_J^2} \biggr)
\simeq  3 \, g_{D^\ast D J \pi}^2  \,.
\en
Based on our Lagrangians we have for the same averaged matrix element squared
\eq\label{Matr1_our}
\sum\limits_{\rm pol} |M_{\rm inv}|^2 &=&
\biggl(\frac{8 g_F g_H}{F_\pi M_J}\biggr)^2  \,
\biggl( M_\pi^2 M_J^2 \biggl( 1 - \frac{M_J^2}{M_Z^2} \biggr)
+ 2 (p_1p_2)^2 \biggl(1 + \frac{M_J^2}{2M_Z^2} \biggr) \, \biggr)
\nonumber\\
&=& \biggl(\frac{8 g_F g_H}{F_\pi M_J}\biggr)^2  \,
\biggl( \frac{\lambda(M_Z^2,M_J^2,M_\pi^2)}{2}
\biggl( 1 + \frac{M_J^2}{2M_Z^2} \biggr) + 3 M_\pi^2 M_J^2
\biggr) \nonumber\\
&\simeq& \biggl(\frac{8 g_F g_H}{F_\pi M_J}\biggr)^2  \,
\frac{(M_Z^2-M_J^2)^2}{2} \,
\biggl(1 + \frac{M_J^2}{2M_Z^2} \biggr) \,.
\en
In above expressions we neglect the pion mass and drop
the loop integral, which is the same in both approaches.

Matching the expressions~(\ref{Matr1_Ko}) and (\ref{Matr1_our})
we derive the constraint on the product of the couplings $g_F g_H$
\eq
g_{D^\ast D J \pi} =
\frac{g_{JDD} \ g_{D^\ast D\pi}}{2 \, \sqrt{2}}  \simeq
\frac{8 g_F g_H }{F_\pi M_J \sqrt{6}} \, (M_Z^2-M_J^2) \,
\sqrt{1 + \frac{M_J^2}{2M_Z^2}} \,.
\en
Here we use the framework of Ref.~\cite{Lin:1999ad} in that the 
$g_{D^\ast D J \pi}$ coupling is expressed
through the product of the $g_{JDD}$ and $g_{D^\ast D\pi}$ couplings.

In complete analogy we derive the relation between the coupling constants
$g_{D^\ast D J \pi}$ and $g_D$ considering the mode $Z_c' \to J/\psi + \pi$:
\eq
g_{D^\ast D^\ast J \pi} \, = \,
\frac{g_{D^\ast D J \pi}}{2 \sqrt{M_D M_{D^\ast}}}
\, = \, \frac{g_{JDD} g_{D^\ast D\pi}}{2 \sqrt{2}} \, = \,
\frac{2 g_D}{F_{\pi}} \,
\frac{M_{Z_c'}^2-M_J^2}{M_{Z_c'}^2}
\en
with $g_{JDD} = 6.5$ fixed in~\cite{Dong:2008gb}, which is a universal constant
for all radially-excited $J(nS)$ states.
One can get a more accurate estimate for these couplings.
We consider the Lagrangian
\eq
{\cal L}_{JDD}(x) = g_{_{JDD}} \, J_\mu(x)
\bar D(x) i \partial^\mu D(x) + {\rm H.c.}
\en
The coupling constant $g_{_{J(nS)DD}}$ is given by
\eq\label{V_universality}
g_{_{J(nS)DD}} = \frac{M_{J(nS)}}{f_{J(nS)}} \,,
\en
where $f_{J(nS)}$ is
determined from the leptonic decays of the $J(nS)$ states as
\eq
\Gamma \Big (\Psi(nS)\to e^+e^-\Big)=
\frac{16\pi\alpha_{_{\rm EM}}^2}{27}\frac{f^2_{J(nS)}}
{M_{J(nS)}}\,,
\en
and $\alpha_{_{\rm EM}} = 1/137.036$ is the fine-structure constant.
The relation (\ref{V_universality}) is the analogue to the $\rho$-meson
universality
\eq
g_{_{\rho\pi\pi}} = \frac{M_\rho}{f_\rho} = \frac{1}{g_{\rho\gamma}}
\en
extended to the heavy quark sector in Ref.~\cite{Lin:2000ke}, where
$g_{\rho\gamma}$ is the $\rho\to\gamma$ transition coupling.

For the last couplings we get
$f_{J(1S)}=416.4$ MeV,
$f_{J(2S)}=295.6$ MeV,
$f_{J(3S)}=187.2$ MeV,
where we used the mass values $M_{J(1s,2s,3s)}=3096.92\pm 0.011~$MeV,
$3686.11 \pm 0.012~$MeV and $4039.6 \pm 4.3~$MeV as well as the results
for the leptonic decay widths of the $J(nS)$ states
\eq
\Gamma\Big (\Psi(1S)\to e^+e^-\Big )&=&5.55\pm 0.14\pm 0.02~\mathrm{keV}\,,
\nonumber\\
\Gamma\Big (\Psi(2S)\to e^+e^-\Big )&=&2.35\pm 0.04~\mathrm{keV}\,,
\nonumber\\
\Gamma\Big (\Psi(3S)\to e^+e^-\Big )&=&0.86\pm 0.07~\mathrm{keV}\, .
\en
Note that we explicitly take into account the $M_{J(nS)}$ dependence of
the $f_{J(nS)}$ and $g_{_{JDD}}$ couplings.
Finally, for the set of $g_{J(nS)DD}$ couplings we get: $g_{_{J(1S)DD}}=7.44$,
$g_{_{J(1S)DD}}=12.47$, $g_{_{J(3S)DD}}=21.58$.

\section{Coupling constants and structure integrals}

The expressions for the coupling constants $g_{_{Z_b}}, g_{_{Z_b'}}$
and structure integrals $J_1$, $J_2$ are
\eq
g_{_{Z_c}}^{-2} &=& \frac{M_{Z_c}^2}{32\pi^2\Lambda^2}
\, \int\limits_0^\infty \frac{d\alpha_1 d\alpha_2}{\Delta_1^3} \,
(\alpha_{12} + 2 \alpha_1 \alpha_2)
\left( 1 + \frac{\Lambda^2}{2M_{D^*}^2\Delta_1} \right) \nonumber\\
&\times&\exp\left\{ -\frac{M_{D^*}^2\alpha_1 + M_{D}^2 \alpha_2}{\Lambda^2}
+ \frac{M_{Z_c}^2}{2\Lambda^2}  \,
\frac{\alpha_{12} + 2 \alpha_1 \alpha_2}{\Delta_1}\right\}\,,\\
g_{_{Z_c'}}^{-2}&=& \frac{M_{Z_c^\prime}^2}{16\pi^2\Lambda^2}
\, \int\limits_0^\infty \frac{d\alpha_1 d\alpha_2}{\Delta_1^2} \,
\left( \frac{\Lambda^2}{M_{Z_c'}^2}
+ \frac{\alpha_{12} + 2 \alpha_1 \alpha_2}{2\Delta_1} \right)
\left( 1 + \frac{\Lambda^2}{M_{D^*}^2\Delta_1} \right) \nonumber\\
&\times&\exp\left\{ -\frac{M_{D^*}^2\alpha_{12}}{\Lambda^2}
+ \frac{M_{Z_c'}^2}{2\Lambda^2}  \,
\frac{\alpha_{12} + 2 \alpha_1 \alpha_2}{\Delta_1}\right\}\,, \\
J_1 &=& \frac{1}{8 \pi^2} \, \int\limits_0^\infty \,
\frac{d\alpha_1 d\alpha_2}{\Delta_2^2} \,
\left( 1+\frac{\Lambda^2}{2M_{D^*}^2 \Delta_2} \right) \nonumber\\
&\times&\exp\left\{ -\frac{M_{D^*}^2\alpha_1 + M_{D}^2 \alpha_2}{\Lambda^2}
+ \frac{M_{Z_c}^2}{4\Lambda^2}  \,
\frac{\alpha_{12} + 4 \alpha_1 \alpha_2}{\Delta_2}\right\}\,,\\
J_2 &=& \frac{1}{8 \, \pi^2} \, \int\limits_0^\infty \,
\frac{d\alpha_1 d\alpha_2}{\Delta_2^2} \,
\left( 1+\frac{\Lambda^2}{M_{D^*}^2 \Delta_2} \right) \nonumber\\
&\times&\exp\left\{ -\frac{M_{D^*}^2\alpha_{12}}{\Lambda^2}
+ \frac{M_{Z_c'}^2}{4\Lambda^2}  \,
\frac{\alpha_{12} + 4 \alpha_1 \alpha_2}{\Delta_2}\right\}\,,
\en
where
\eq
\Delta_1=2 + \alpha_{12} \,, \quad \Delta_2=1 + \alpha_{12} \,, \quad
\alpha_{12} = \alpha_1 + \alpha_2 \,.
\en

\newpage 
\begin{figure}
\centering
\includegraphics [scale=0.5]{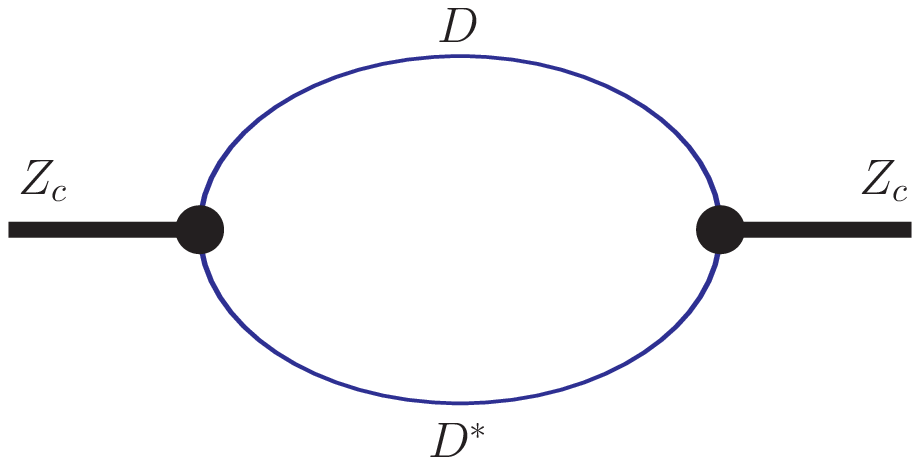}
\hspace{1cm}
\includegraphics [scale=0.5]{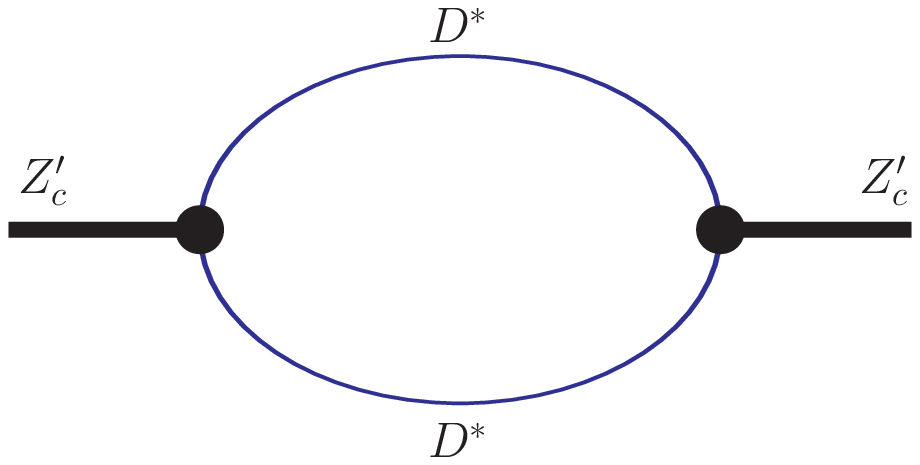}
\caption{Mass operators of $\zc$ and $\zcp$.}

\vspace*{2.5cm}
\includegraphics [scale=0.6]{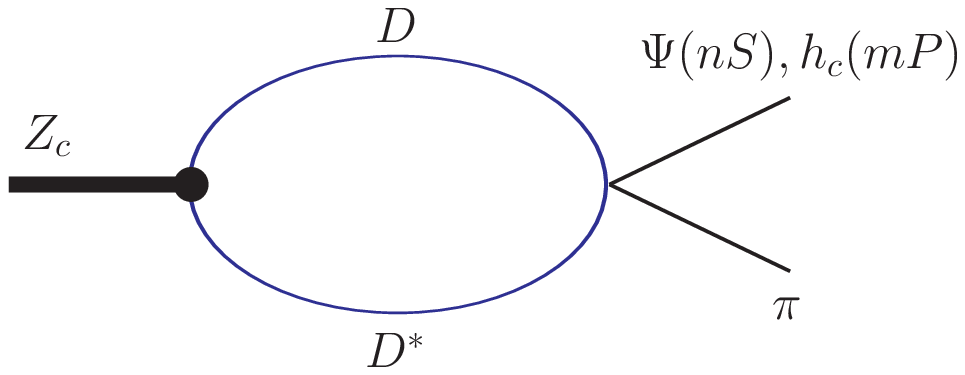}
\hspace*{.5cm}
\includegraphics [scale=0.6]{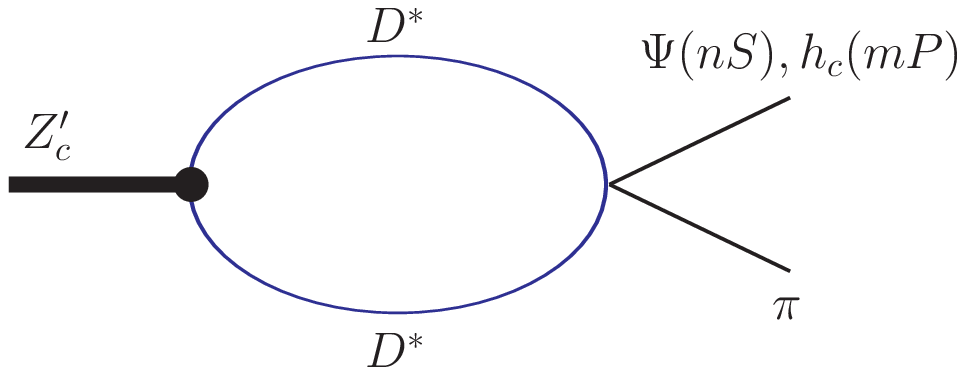}
\caption{Two-body decays $\zc \to \Psi(nS),h_c(mP) + \pi$ and
$\zcp \to \Psi(nS),h_c(mP) + \pi$.}
\end{figure}

\vspace*{1cm}

\begin{center}
{\bf Table I.} Phenomenological couplings $g_{_{Z_cH\pi}}$ and
$g_{_{Z_c'H\pi}}$ in GeV$^{-1}$.

\vspace*{.25cm}
\def\arraystretch{1.2} 
\begin{tabular}{|c|c|c|c|c|c|c|}\hline
$\epsilon$ &$g_{_{Z_c\Psi(1S)\pi}}$ &$g_{_{Z'_c\Psi(1S)\pi}}$
&$g_{_{Z_c\Psi(2S)\pi}}$  &$g_{_{Z'_c\Psi(2S)\pi}}$  &$g_{_{Z_ch_c(1P)\pi}}$
&$g_{_{Z'_ch_c(1P)\pi}}$  \\ \hline
5  &0.81-1.10 &0.83-1.26 &4.78-6.47 &3.60-6.47 &0.81-1.10 &0.68-1.03\\ \hline
10 &0.88-1.31 &0.96-1.44 &5.27-7.89 &4.23-7.89 &0.88-1.31 &0.79-1.18\\ \hline
15 &0.94-1.41 &1.05-1.58 &5.75-8.66 &4.65-8.66 &0.94-1.41 &0.86-1.29\\ \hline
20 &0.99-1.49 &1.10-1.68 &6.20-9.38 &4.95-9.39 &0.99-1.49 &0.90-1.37\\ \hline
\end{tabular}

\vspace*{.5cm}

{\bf Table II.} Values of K\"allen functions for 
different binding energies in GeV$^{4}$ for $M_\pi \equiv M_{\pi^+}$. 

\vspace*{.25cm}
\def\arraystretch{1.2}
\begin{tabular}{|c|c|c|c|c|c|c|}\hline
$\epsilon$ (MeV) &$\lambda_{_{Z_c\Psi(1S)\pi}}$ &$\lambda_{_{Z'_c\Psi(1S)\pi}}$
&$\lambda_{_{Z_c\Psi(2S)\pi}}$   &$\lambda_{_{Z'_c\Psi(2S)\pi}}$
&$\lambda_{_{Z_ch_c(1P)\pi}}$ &$\lambda_{_{Z'_ch_c(1P)\pi}}$   \\ \hline
5  &28.183 &41.414  &0.848 &5.157  &5.470 &12.364\\ \hline
10 &27.767 &40.895  &0.741 &4.959  &5.275 &12.074\\ \hline
15 &27.358 &40.380  &0.638 &4.765  &5.084 &11.786\\ \hline
20 &26.950 &39.869  &0.540 &4.573  &4.895 &11.502\\ \hline
\end{tabular}

\vspace*{.5cm}
{\bf Table III.} Predictions for the strong decay widths of $\zc$ and
$\zcp$ states in MeV.

\vspace*{.25cm}
\def\arraystretch{1.2}
\begin{tabular}{|c|c|c|c|c|c|c|}\hline
 $\epsilon$ (MeV) 
&$\Gamma_{Z_c}(1S) $ 
&$\Gamma_{Z_c'}(1S)$
&$\Gamma_{Z_c}(2S) $  
&$\Gamma_{Z_c'}(2S)$  
&$\Gamma_{Z_c}(1P) $
&$\Gamma_{Z_c'}(1P)$  \\ \hline
5  &7.45-13.63  &11.50-26.60 &1.47-2.70 &8.26-19.1   &0.68-1.25 &1.02-2.36\\ 
\hline
10 &8.53-19.15  &15.33-34.39 &1.48-3.32 &10.81-24.23 &0.76-1.70 &1.34-3.00\\ 
\hline
15 &9.55-21.66  &17.85-40.64 &1.41-3.21 &12.32-28.06 &0.82-1.86 &1.53-3.49\\ 
\hline
20 &10.43-23.89 &19.47-45.11 &1.28-2.94 &13.16-30.48 &0.87-1.98 &1.65-3.81\\ 
\hline
\end{tabular}
\end{center}

\end{document}